# A Calculation of the Sigma Meson Mass in Nuclear Matter

J. R. Morones-Ibarra[1] and Ayax Santos-Guevara[1]
(1) Facultad De Ciencias Físico-Matemáticas, Universidad Autónoma de Nuevo León, Ciudad Universitaria, San Nicolás de los Garza, Nuevo León, 66450, México.

**Abstract**

We calculate the mass and width of the sigma meson in nuclear medium by considering that it couples to two virtual pions and to a pair of nucleon-antinucleon states and to particle-hole states. The mass is calculated by using the spectral function in the Walecka model, finding that it is about $380 MeV$. In addition, we obtain the value of $340 MeV$ for the width of its spectral function. We find that there is a reduction in the mass value compared with that in vacuum. This result is consistent with those reported by other authors who have used different models predicting a decreasing of the mass as a function of the density.

## 1. Introduction

The study of the properties and structure in vacuum and in nuclear matter of the scalar-isoscalar $\sigma$ meson has recently attracted growing attention because of the role that this particle plays in some theoretical models and the increasing experimental evidence of its existence [1,2]. It has been assumed that the $\sigma$ meson participates as an intermediate particle in several processes in vacuum and in hot and dense nuclear matter [3]. Particularly, this meson is of wide interest because in theoretical models of the nuclear force, the sigma is the responsible for the attractive part of the nuclear potential [4], which is attributed to the exchange of a single $\sigma$-meson.

The sigma meson also plays the role of the chiral partner of the pion in the sigma model, which is a toy model for the interaction between nucleons and pions, with $SU(2)_L \otimes SU(2)_R$ symmetry [5,6,7,8]. This model was originally developed by M. Gell-Mann and M. Levy [9], In this model the σ- meson is analogous to the Higgs particle in the Weinberg-Salam theory [7,9], in the sense that the nucleon gets mass when the $SU(2)_L \otimes SU(2)_R$ symmetry is spontaneously broken. In addition, in chiral perturbation theory the sigma meson enters as an essential part to adjust the theory to the experimental data [10], and in the interactions between mesons, the $\sigma$ appears as a resonance, showing up as a pole in the $T$ matrix [11, 12]. In [6, 13, 14] they propose some experimental possibilities for investigating the behaviour of the sigma meson in hot and in dense nuclear matter.

However, the existence and the composition of the $\sigma$-meson, as a $\bar{q}q$ meson or as a $\pi\pi$ resonance [13,15,16], is a subject of controversy . We concluded in a recent work that it can be considered as a two pion resonance [17]

For describing the nuclear matter we use in this work the Walecka model, which is a renormalizable relativistic quantum field theory where the degrees of freedom are nucleons that interact through the exchange of scalar meson, the sigma, and vector mesons, the omega.

In this work we study the scalar sigma meson in nuclear medium when the meson couples to two nucleons through a Yukawa-type coupling. The interaction of the meson with nucleons includes interactions with the Fermi Sea producing nucleon-hole states and interactions with the Dirac Sea producing nucleon-antinucleon states [4, 19]. The interaction of sigma with two virtual pions is also included in the calculations.

We define the mass and width in terms of the spectral function. Our definition of mass for a particle is the magnitude $|k|$ of its four-momentum for which the particle spectral function $S(k^2)$ gets its maximum. In this work we find a closed expression for the regularized self-energy function of the σ- meson and then we obtain an exact analytical function for its spectral function. The σ- meson self- energy is calculated in the one loop level and the propagator is computed by summing over ring diagrams, in the so called Random Phase Approximation (RPA) [18]. To carry out the summation we use the Dyson's equation. The real part of the self-energy in the pion-sigma interaction is ultraviolet divergent and it is regularized by using a simple subtraction dispersion relation, which preserves the symmetries of the theory.

## 2. Formalism

There are several ways of defining and determining theoretically the mass and the width of unstable particles. Some authors, as in [5,9,10, 24] make use of the spectral function to define a mass of a particle. In [26,27] the S-matrix formalism is used to determine the mass and the width of a meson. On the other hand, authors as in [8,13,14,15], the mass of a particle is defined as the pole in its complete propagator.

The definition of the mass of a particle in terms of its spectral function, is used extensively in the literature, [24,28,29] and it is well established. This is the definition that we will use in this work for calculating the $\sigma$-meson mass in vacuum and in nuclear matter.

In order to evaluate the mass of the σ- meson we need, firstly, to calculate its dressed propagator $i\Delta(k)$ where $k$ is the four-momentum of the propagating meson. The expression for the dressed σ- meson propagator $i\Delta(k)$ is obtained from the Dyson equation, [4,19]

$$i\Delta(k) = i\Delta_0(k) + i\Delta_0(k)[-i\Sigma(k)]i\Delta(k) \qquad (1)$$

where $\quad i\Delta_0(k) = \dfrac{i}{k^2 - \left(m_\sigma^0\right)^2 + i\varepsilon}$

is the free σ- meson propagator, with $m_\sigma^0$ and $\Sigma(k)$ being the bare mass and the self-energy of σ, respectively. The self-energy $\Sigma(k)$ contains all the information about the interactions of the meson with the quantum vacuum and nuclear matter, so, in order to determine the self energy $\Sigma(k)$ we must specify the dynamical content of our model.

The Lagrangian density for the Walecka model [4] is

$$L = \overline{\Psi}\left[\gamma_\mu\left(i\partial^\mu - g_v V^\mu\right) - \left(M - g_s \Phi\right)\right]\Psi + \frac{1}{2}\left(\partial_\mu \Phi \partial^\mu \Phi - (m_\sigma^0)^2 \Phi^2\right) - \frac{1}{4}F_{\mu\nu}F^{\mu\nu}$$
$$+ \frac{1}{2}m_v^2 V_\mu V^\mu + \delta L \quad (2)$$

where $\Psi$ is the nucleon field with mass $M$, $V^\mu$ is the neutral vector meson field ($\omega$) with mass $m_v$, $F^{\mu\nu} = \partial^\mu V^\nu - \partial^\nu V^\mu$ is the tensor field of the vector meson $\omega$, $\Phi$ is the scalar ($\sigma$) meson field with the bare mass $m_\sigma^0$; $g_s$ and $g_v$ are the coupling constants, and finally $\delta L$ contains renormalization counterterms.

The interaction Lagrangian density $L_{\sigma\pi\pi}$ which describes the σ-π dynamics, [30] is given by

$$L_{\sigma\pi\pi} = \frac{1}{2}g_{\sigma\pi\pi}m_\pi \vec{\pi} \cdot \vec{\pi} \Phi \quad (3)$$

where $\vec{\pi} = (\pi_1, \pi_2, \pi_3)$ represents the Cartesian components of the pseudoscalar π-meson field, $\Phi$ is the scalar σ-meson field, $g_{\sigma\pi\pi}$ is the coupling constant, and $m_\pi$ is the mass of the π-meson.

The influence of the interaction of σ-mesons with virtual pions and nucleons in nuclear matter is introduced through the modification of the free propagator in the one loop approximation; this is shown graphically in the Fig. 1. The dashed lines represent the σ-meson, the dotted lines represent the pion field and the continuum lines are associated to the field of nucleons. We will calculate the full propagator in the chain approximation, which consists in an infinite summation of the one loop self-energy diagrams [22]. The diagrammatical representation of the modified propagator is showed in Fig. 2, and the analytical expression is given by $i\Delta(k)$ in Eq. (1).

The solution for $\Delta(k)$ in the Dyson's equation (1) is given by

$$\Delta(k) = \frac{1}{\left[\Delta_0(k)\right]^{-1} - \Sigma(k)} = \frac{1}{k^2 - \left(m_\sigma^0\right)^2 - \Sigma(k)} \quad (4)$$

On the other hand, the analytical expression for the self-energy $\Sigma(k)$, is given by [5,30]

$$\Sigma(k) = \Sigma_{\sigma\pi}(k) + \Sigma_{\sigma N}(k),$$

where

$$-i\Sigma_{\sigma\pi}(k) = \frac{3}{2}g_{\sigma\pi\pi}^2 m_\pi^2 \int \frac{d^4q}{(2\pi)^4} \frac{1}{q^2 - m_\pi^2 + i\varepsilon} \frac{1}{(q-k)^2 - m_\pi^2 + i\varepsilon} \qquad (5)$$

$$i\Sigma_{\sigma N}(k) = -ig_s^2 \int \frac{d^4q}{(2\pi)^4} Tr[G(q)G(q+k)] \qquad (6)$$

where the coefficient 3/2 for the pion loop comes from the three isospin states and the permutation symmetry factor [30], and $G(q)$ is the full nucleon propagator [4]

$$G(q) = (\gamma_\mu q^{*\mu} + M)\left[\frac{1}{q^{*2} - M^{*2} + i\varepsilon} + \frac{i\pi}{E(q)}\delta(q_0 - E(q))\Theta(k_F - |\vec{q}|)\right]$$

$$G(q) = G_F(q) + G_D(q) \qquad (7)$$

where $q^{*\mu} \equiv (q^0 - g_v V^0, \vec{q})$, $E^*(q) \equiv \sqrt{\vec{q}^2 + M^{*2}}$, $k_F$ is the Fermi momentum, and $M^*$ is the nucleon effective mass.

### 2.1 Calculation of $\Sigma(k)$ in vacuum

Carrying out the integration in Eq.. (5) respect to $q_0$ by using the Cauchy residue theorem, integrating in the $q_0$ complex plane, we obtain

$$\Sigma(k) = -\frac{3}{8\pi^2}g_{\sigma\pi\pi}^2 m_\pi^2 \int_{-\infty}^{\infty} \frac{\vec{q}^2 d|\vec{q}|}{\sqrt{(\vec{q}^2 + m_\pi^2)}\left[4(\vec{q}^2 + m_\pi^2) - k_0^2 - i\varepsilon\right]} \qquad (8)$$

Separating this expression in the real and imaginary parts and integrating the imaginary part, we obtain

$$\text{Im}\Sigma_{\sigma\pi}(k^2) = \frac{-3g_{\sigma\pi\pi}^2}{32\pi}m_\pi^2\left(1 - \frac{4m_\pi^2}{k^2}\right)^{1/2} \qquad (9)$$

for $k^2 \geq 4m_\pi^2$, and zero for $k^2 < 4m_\pi^2$. We can see the characteristic threshold value $k^2 \geq 4m_\pi^2$ for the production of real $\pi - \pi$ pairs from the σ-field.

On the other hand, the real part of $\Sigma_{\sigma\pi}(k^2)$ is ultraviolet divergent, and therefore it needs to be regularized. The regularization of $\text{Re}\Sigma_{\sigma\pi}(k^2)$ will be done by using a simple subtraction dispersion relation [31], which is given by the expression

$$\Sigma(t) = \frac{t}{\pi}\int_0^{\infty} \frac{\text{Im}\Sigma_{\sigma\pi}(t')}{t'(t'-t) - i\varepsilon}dt' \qquad (10)$$

For the imaginary part in the integrand taking from Eq. (9), the real part of the integral in Eq. (10) is convergent [31].

Now let us write the identity

$$\operatorname{Re}\Sigma_{\sigma\pi}(k^2) = \operatorname{Re}\Sigma_{\sigma\pi}(k^2) - \operatorname{Re}\Sigma^0_{\sigma\pi}(k^2) + \operatorname{Re}\Sigma^0_{\sigma\pi}(k^2) \qquad (11)$$

where $\operatorname{Re}\Sigma^0_{\sigma\pi}(k^2)$ is an infinite quantity chosen conveniently to cancel the infinite terms of $\operatorname{Re}\Sigma_{\sigma\pi}(k^2)$. We define now the finite quantity $\operatorname{Re}\Sigma^R_{\sigma\pi}(k^2) = \operatorname{Re}\Sigma_{\sigma\pi}(k^2) - \operatorname{Re}\Sigma^0_{\sigma\pi}(k^2)$ as the regularized real part of the σ-meson self energy, and the renormalized mass $m_\sigma$ through $m_\sigma^2 = (m_\sigma^0)^2 + \operatorname{Re}\Sigma^0_{\sigma\pi}(k^2)$. With this definition, $m_\sigma$ is taking as the experimental value of the mass for sigma.

From Eq. (10), we obtain for the real part

$$\operatorname{Re}\Sigma^R_{\sigma\pi}(k^2) = -\frac{3m_\pi^2 g_{\sigma\pi\pi}^2 k^2}{32\pi^2} P\int_{4m_\pi^2}^{\infty} \frac{(1-\frac{4m_\pi^2}{x'})^{\frac{1}{2}}}{x'(x'-k_0^2)}dx' \qquad (12)$$

This integral can be carried out directly giving the result

$$\operatorname{Re}\Sigma^R_{\sigma\pi}(k^2) = -\frac{3g_{\sigma\pi\pi}^2 m_\pi^2}{16\pi^2}(1+cI_0), \qquad (13)$$

where

$$c \equiv 1 - \frac{4m_\pi^2}{k^2} \text{ and,}$$

$$I_0 = \frac{1}{2\sqrt{c}}\ln\left|\frac{\sqrt{c}-1}{\sqrt{c}+1}\right| \text{ with } c > 0$$

The expression for the renormalized self-energy $\Sigma^R_{\sigma\pi}(k^2) = \operatorname{Re}\Sigma^R_{\sigma\pi}(k^2) + i\operatorname{Im}\Sigma_{\sigma\pi}(k^2)$, constructed from Eqs. (9) and (13), is part of the main result of this work.

## 2.2 Calculation of Σ(k) in nuclear matter

### 2.2.1 Vacuum fluctuations (Dirac sea contribution)

The term $G_F(q)$ in the nucleon propagator, which is a divergent quantity, contains the nucleon-antinucleon contribution in the propagator $G(q)$. Subtracting the appropriate counterterms $CTC$ [32] we made this term finite

$$\Sigma^R_{\sigma N}(k) = \Sigma^{RF}_{\sigma N}(k) + \Sigma^D_{\sigma N}(k) \qquad (14)$$

$$\Sigma_{\sigma N}^{RF}(k) = \Sigma_{\sigma N}^{F}(k) - CTC \tag{15}$$

In Eqs. 14-15 superscript $R$ means renormalized.

Taking in Eq. (6) the terms that do not depend on the density, we obtain

$$\Sigma_{\sigma N}^{F}(k) = -8ig_s^2 \int \frac{d^4q}{(2\pi)^4} \frac{q^2 + q \cdot k + M^{*2}}{[q^2 - M^{*2} + i\varepsilon][(q+k)^2 - M^{*2} + i\varepsilon]} \tag{16}$$

Now we use in Eq. (16) the Feynman parametrization technique and then the dimensional regularization method [33] to isolate the divergent terms. After some algebra we get to

$$\Sigma_{\sigma N}^{F}(k) = \frac{6g_s^2}{4\pi^2}\left(\frac{2}{\varepsilon} - \gamma_E\right)\int_0^1 dz [M^{*2} - k^2 z(1-z)][1 - \frac{\varepsilon}{2}\ln(M^{*2} - k^2 z(1-z))] \tag{17}$$

With the purpose to make Eq. (17) finite, we subtract four counterterms [4,5,32]. Each counterterm is evaluated in vacuum, where $M^* = M$ and $k_F = 0$. Besides, each counterterm is evaluated in an appropriate value of $q^2$ corresponding to our renormalization point. Our choice is the same as in [32]:

$$k^2 = 0 \tag{18}$$

Then, Eq. (17) can be written as

$$\Sigma_{\sigma N}^{F}(k) = \frac{3g_s^2}{2\pi^2}\left\{M^2 + M^{*2} - 4MM^* - \frac{k^2}{6} + \gamma_E(2M^{*2} - 2M^2) - M^{*2}\ln\frac{M^{*2}}{M^2} + \frac{k^2}{6}\ln M^{*2}\right.$$

$$\left. - \frac{4}{9}k^2 + \frac{k^2}{6}a^2 - a\left[2M^{*2} + \frac{1}{2} - \frac{1}{6}a\right]\int_0^1 \frac{dv}{v^2 - a}\right\} \tag{19}$$

with $a \equiv 1 - \frac{4M^{*2}}{k^2}$, $\gamma_E$ being the Euler constant and the integral

$$\int_0^1 \frac{dv}{v^2 - a} = \begin{cases} \dfrac{1}{2\sqrt{a}} \ln \dfrac{\sqrt{a}-1}{\sqrt{a}+1}, & \text{for } k^2 < 0 \\ \dfrac{1}{\sqrt{-a}} \arctan\left(\dfrac{1}{\sqrt{-a}}\right), & \text{for } 0 < k^2 < 4M^{*2} \\ \dfrac{1}{2\sqrt{a}} \ln \dfrac{\sqrt{a}-1}{\sqrt{a}+1} + \dfrac{i\pi}{2\sqrt{a}}, & \text{for } k^2 > 4M^{*2} \end{cases}$$

### 2.2.2 Density-dependent term

Now we calculate the density dependent term of Eq. (6)

$$\Sigma_{\sigma N}^D(k) = -8ig_s^2 \int \frac{d^4q}{(2\pi)^4}(q^2 + q \cdot k + M^{*2}) \left\{ \frac{\delta[(q^0 - k^0) - E(q+k)]\Theta(k_F - |\vec{q}+\vec{k}|)}{[q^2 - M^{*2}]E(q+k)} \right.$$

$$\left. + \frac{\delta[(q^0 - k^0) - E(q)]\Theta(k_F - |\vec{q}|)}{[(q+k)^2 - M^{*2}]E(q)} \right\}. \quad (20)$$

Carrying out the integration of $\Sigma_{\sigma N}^D(k)$ respect to $q^0$ we obtain

$$\Sigma_{\sigma N}^D(k) = -8ig_s^2 \int \frac{d^3q}{(2\pi)^4} \left\{ \frac{\Theta(k_F - |\vec{q}|)[(q-k)^2 + (q-k) \cdot k + M^{*2}]}{[-2q \cdot k + k^2]E(q)} \right.$$

$$\left. + \frac{\Theta(k_F - |\vec{q}|)}{[2q \cdot k + k^2]E(q)} \right\} \quad (21)$$

Simplifying Eq. (23), and taking the condition $(k^2)^2 \ll 4(q \cdot k)^2$ for low energies [19], we find

$$\Sigma_{\sigma N}^D(k) = -8ig_s^2 \int \frac{d^3q}{(2\pi)^4} \frac{[k^2 - (q-k) - M^{*2}]}{[-4(q \cdot k)^2]E(q)} \Theta(k_F - |\vec{q}|) \quad (22)$$

Taking $\vec{k}$ in the positive direction of the $z$ axis, this means $k^\mu = (k^0, 0, 0, k^{(3)})$, $E(\vec{q}) \equiv \sqrt{\vec{q}^2 + M^{*2}}$, and defining $\chi \equiv -sen\theta d\theta$, we obtain, after integration, the expression

$$\Sigma_{\sigma N}^D(k) = -\frac{2g_s^2 M^{*2}}{\pi^2} \left\{ \ln \frac{k_F + \varepsilon_F}{M^*} + \frac{C_0}{2} \ln \left| \frac{(k_F - A)(\varepsilon_F \sqrt{A^2 + M^{*2}} - Ak_F + M^{*2})}{(k_F + A)(\varepsilon_F \sqrt{A^2 + M^{*2}} + Ak_F + M^{*2})} \right| \right\}$$

$$+ \frac{g_s^2}{\pi^2}\left\{k_F \varepsilon_F - M^{*2} \ln \frac{k_F + \varepsilon_F}{M^*}\right\} \qquad (23)$$

where $C_0 \equiv \frac{k^0}{|\vec{k}|}$ and $A \equiv \frac{C_0 M^*}{\sqrt{1-C_0^2}}$.

## 3. Results

The propagator given by Eq. (4), takes the form

$$\Delta(k^2) = \frac{1}{k_0^2 - m_\sigma^2 - \mathrm{Re}\Sigma^R(k^2) - i\,\mathrm{Im}\Sigma(k^2)} \qquad (24)$$

where

$$\Sigma(k) = \Sigma_{\sigma\pi}(k) + \Sigma_{\sigma N}(k) \qquad (25)$$

and

$$\mathrm{Re}\Sigma^R(k) = \mathrm{Re}\Sigma_{\sigma\pi}(k) + \Sigma_{\sigma N}(k)$$
$$\mathrm{Im}\Sigma(k) = \mathrm{Im}\Sigma_{\sigma\pi}(k),$$

being $\Sigma_{\sigma N}(k)$ the sum of Eqs (19) and (23).

From the definition of the spectral function $S(k^2)$ given above, we have

$$S(k^2) = -\frac{2\pi\,\mathrm{Im}\Sigma(k^2)}{\left[k_0^2 - m_\sigma^2 - \mathrm{Re}\Sigma^R(k^2)\right]^2 + \left[\mathrm{Im}\Sigma(k^2)\right]^2} \qquad (26)$$

Substituting $\mathrm{Im}\Sigma(k^2)$ and $\mathrm{Re}\Sigma^R(k^2)$ from Eqs.(9), (13), (19) and (23) into Eq. (26), we obtain a closed expression for the spectral function. The parameters in Eq. (26) are the reported $m_\sigma$ value for the σ-meson mass, which we take as $700 MeV$, and the bare σππ coupling constant $g_{\sigma\pi\pi} = 12.8$ [30].

The spectral function of the sigma meson has been plotted in Fig. 3 at normal nuclear matter density. The nucleon mass was fixed at their physical value ($M = 939 MeV$). The effective nucleon mass $M^*$ is the appropriate value at nuclear-matter saturation density. This relation was taken as $M^*/M = 0.730$, and the bare σN coupling constant $g_s^2 = 54.289$ [22].

**Conclusions**

As we can see, the maximum of $S(k^2)$ is getting at $k = 380 MeV$ at normal nuclear matter density. This value is in agreement with that reported by other authors, using different models as the Nambu Jona-Lasinio [34] and the chiral perturbation theory [11,12], which predict a decreasing of the mass of the sigma meson when it is in nuclear matter. This result is interpreted as a partial restoration of chiral symmetry [35,36]. On the other hand, we obtained the value of $340 MeV$ for the width, taking this at one half of the maximum value of the spectral function.

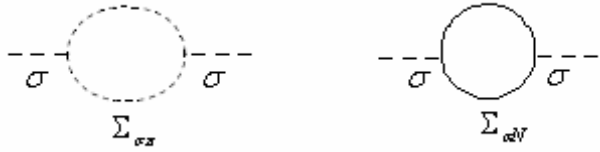

Fig. 1. Sigma meson self-energy diagram. The dashed lines represent the σ-meson, the dotted lines the pion field and the continuous line the nucleon field.

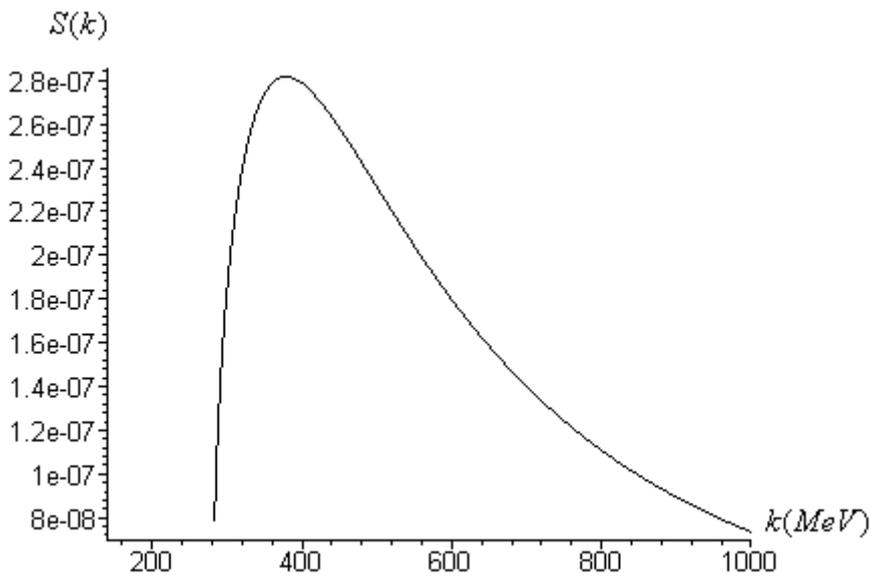

Fig. 2 Sigma meson full propagator in the chain approximation. The left hand side represents the dressed sigma propagator and the right hand side is the diagrammatical representation of the chain approximation to the dressed propagator.

Fig. 3 Spectral function of sigma meson in nuclear matter.